\documentclass[prl,twocolumn,noshowkeys,superscriptaddress,preprintnumbers,amsmath,amssymb, showpacs]{revtex4}

\newcommand{\trace}[1]{\textsf{tr}(#1)}

\newcommand{\braket}[2]{\left\langle #1 \vert #2\right\rangle}
\newcommand{\abs}[1]{\left\vert #1 \right\vert}
\newcommand{\Real}[1]{\textsf{Re}(#1)}
\newcommand{\Imag}[1]{\textsf{Im}(#1)}

\usepackage{amsmath}
\usepackage{bbm,amsfonts}

\def\E{{\mathcal E}}

\usepackage{amsthm}
\theoremstyle{definition}

\input{Qcircuit.tex}

\begin{document}

\title{Selective Efficient Quantum Process Tomography}

\author{Ariel Bendersky}
\affiliation{Departamento de F\'\i sica, FCEyN UBA, Pabell\'on 1, Ciudad Universitaria, 1428 Buenos Aire, Argentina}
\author{Fernando Pastawski}
\affiliation{FaMAF, Universidad Nacional de C\'ordoba and Universidad de Buenos Aires}
\author{Juan Pablo Paz}
\affiliation{Departamento de F\'\i sica, FCEyN UBA, Pabell\'on 1, Ciudad Universitaria, 1428 Buenos Aire, Argentina}
\date{\today}

\begin{abstract} We present a new method for quantum process tomography. The method enables us to efficiently estimate, with fixed precision, any of the parameters characterizing a quantum channel. It is selective since one can choose to estimate the value of any specific set of matrix elements of the super-operator describing the channel. Also, we show how to efficiently estimate
all the average survival probabilities associated with the channel (i.e., all the 
diagonal elements of its $\chi$--matrix). 
\end{abstract}


\maketitle

The efficient characterization of the temporal evolution of a quantum system is one of the main tasks one needs to accomplish to achieve quantum information processing. 
In particular, this is essential to determine the most important sources of errors to design appropriate quantum error correction techniques.
The set of methods used to determine the evolution of a quantum system are generically denoted as  {quantum process tomography} (QPT). 
In general, QPT is a very hard task whose completion typically requires resources scaling exponentially with the number of qubits in the system ($n$). 
To show that this is indeed the case, we can proceed as follows: 
Under very general assumptions the evolution of the quantum state of a system can be represented by a linear, completely positive, trace preserving map $\rho_{out}=\E(\rho_{in})$. 
Choosing a base of $D^2$ operators $\{E_m\}$, the map can be written as:
\begin{equation}
\E(\rho) = \sum_{mn} \chi_{mn} E_m \rho E_n^\dagger\ 
~\text{ with }~
\sum_{mn} \chi_{mn} E_n^\dagger E_m = I.
\label{eq:tracecond} 
\end{equation}
Thus, the map is completely characterized by the positive hermitian matrix $\chi_{mn}$ that must also satisfy the above trace perserving condition. Therefore, fully characterizing the map requires $D^4-D^2$ real parameters (where $D=2^n$ is the dimension of the Hilbert space of $n$ qubits). For this reason QPT is a highly inefficient task. 

In this letter we present a method that enables us to evaluate (with fixed precision) any of the coefficients of the $\chi$--matrix with resources that scale polynomially with the number of qubits. 
The method is ``selective'' since one can use it to estimate any coefficient (or any set of coefficients). 
For each coefficient there is an efficient estimation strategy that we describe below. 
For this reason we denote our strategy as  \emph{selective efficient quantum process tomography} (SEQPT).
Our method is inspired on previous proposals that use randomized subroutines as 
intermediate steps for efficiently estimating any average gate fidelity.
Moreover, we also propose a method to estimate all the 
diagonal $\chi_{mm}$ coefficients using only polynomial resources.

First, it is convenient to review the main tomographic schemes and their properties. 
{Standard quantum process tomography} (SQPT) was the first tomographic method proposed \cite{NielsenChuang00}. 
It involves preparing a set of input states $\rho_k$, and then performing full quantum state tomography on the resulting output states obtained after evolution. 
By doing this, we directly measure coefficients $\lambda_{jk}=Tr(\rho_k\E(\rho_j))$. 
However, if one wants to find the matrix elements $\chi_{mn}$ it is necessary to invert an exponentially large system of equations relating $\lambda$ with $\chi$ \cite{chuang97prescription}. 
For this reason, the method is indirect (since it requires inversion to obtain matrix elements $\chi_{mn}$). It is also inefficient since, in the most general case, in order to estimate any of the coefficients $\chi_{mn}$ one needs to perform an exponentially large number of experiments and classical postprocessing. 

{Direct Characterization of Quantum Dynamics} (DCQD) was recently proposed
\cite{mohseni-2006-97, mohseni-2007-75} and it requires an extra ancillary system of $n$ clean qubits with a clean quantum channel. 
Provided such a resource is available, 
the method enables the direct estimation of all diagonal $\chi_{mm}$ 
by associating each result of a single complete measurement to each one.
The method requires a rather small number of elementary quantum gates ($O(n)$).
A prescription for estimating any $\chi_{mn}$ is given. 
However, estimating off-diagonal coefficients may require 
the inversion of a system of equations which, for some off-diagonal $\chi_{mn}$  coefficients, involves an exponential number of experiments.
In these cases this method is inefficient.

{Symmetrized Characterization of Noisy Quantum Processes} \cite{JosephEmerson09282007} (SCNQP) transforms the channel $\E$ into a symmetrized channel $\E'$ via twirling operations.
After symmetrization, only diagonal $\chi'_{mm}$ coefficients remain, being the averages over the original coefficients of the same Hamming weight.
The twirling is achieved using only ($O(n)$) single qubit gates with constant depth.
The values of the averaged coefficients are linearly related to output probabilities through an upper diagonal square matrix of size $n+1$.
The method is ideally tailored for evaluating the applicability of relevant quantum error correcting codes \cite{2007arXiv0710.1900S}. 
as it allows the evaluation of diagonal $\chi_{mm}$ coefficients averaged over operators of the same Hamming weight (i.e. $\chi_{00}$, average over $1$ qubit errors, etc).
However, it is not possible to estimate any of the off-diagonal $\chi_{mn}$ coefficients, which are wiped out by the symmetrization protocol, nor distinguish among specific Pauli errors of the same Hamming weight.

Our method has a similar flavor to SCNQP adding the possibility to determine any of the coefficients $\chi_{mn}$ (including off-diagonal ones) with polynomial resources. The method is based on {two observations}: 
The first, is the fact that any matrix element $\chi_{mn}$ can be related to an average survival probability of input states under the action of the channel (or a related quantity as described below). 
The  average involved here is over the entire Hilbert space using the so-called Haar measure. 
The second observation is that such averages can be efficiently estimated by sampling over a finite set of states (a $2$--design, as described below). 
Let us describe these two crucial observations in more detail. Before doing that we point out that a choice of base to define the $\chi_{mn}$ matrix is required. We will use an operator base $\{E_m\}$ satisfying $\trace{E_m E_n^\dagger}=D\delta_{m,n}$ (orthonormality) and. $E_m E_m^\dagger=I$ (unitarity). 
For definiteness only, we assume $E_0 = I$. A convenient choice is the base formed by tensor product of Pauli operators $\{P_m\}$, which have a convenient group structure: $P_m P_n = i^{\theta_{m,n}}P_{m\star n}$, where $\star$ is the commutative group operation.

We start by noticing that the average fidelity of the map $\E$ is defined as: \cite{nielsen-2002-303}
\begin{equation}
F(\E)=\int \bra{\psi}\E(\ket{\psi}\bra{\psi})\ket{\psi} d \psi 
\label{eq:avgfidelity} 
\end{equation}
In fact, $F(\E)$ is nothing but the survival probability averaged over all pure states $\ket{\psi}$. It is quite simple to see that there is a direct relation between the average fidelity $F(\E)$ and the coefficient $\chi_{00}$. 
As was shown in \cite{renes-2004-45}, for any operators $O_1$ and $O_2$ we have:
\begin{equation}
\int \bra{\psi}O_1 P_\psi O_2 \ket{\psi} d\psi = \frac{\trace{O_1}\trace{O_2}+\trace{O_1 O_2}}{D(D+1)}.  
\label{eq:avgastraces}
\end{equation}
where $P_\psi=\ket{\psi}\bra{\psi}$ is the projector on state $\ket{\psi}$.
Using this equation together with the $\chi$--matrix representation for the map $\E$ in the definition of the average fidelity (\ref{eq:avgfidelity}) we find $ F(\E) = \frac{D \chi_{00} + 1}{D+1}$. 
All other diagonal coefficients $\chi_{mm}$ are directly related to average fidelities of slightly modified channels. 
In fact, if we consider the channel $\E_m(\rho)=E^\dagger_m\E(\rho)E_m$, its fidelity is given by:
\begin{equation}
F(\E_m) = \int \bra{\psi}E_m^\dagger \E(P_\psi)E_m\ket{\psi} d\psi = \frac{D \chi_{mm}+1}{D+1}
\label{eq:avggatefidelity}
\end{equation}
Note that all the above fidelities have a lower bound of $1/(D+1)$.
To measure the diagonal coefficients $\chi_{mm}$ we must 
perform the experiment described in Fig. 1 and average over all states $\ket{\psi}$. 
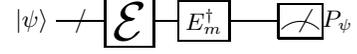
\begin{figure}[h]
\begin{equation*}
\Qcircuit @C=1em @R=.7em {
     \ket{\psi} & &  {/} \qw &\gate{\mbox{\huge $\E$}} & \gate{E_m^\dagger} &\qw & \meter   
}
P_\psi 
\end{equation*}
\caption{Circuit for measuring $\chi_{mm}$ for a given channel $\E$}
\label{circ:diag}
\end{figure}

The off-diagonal elements $\chi_{mn}$ can also be related to average quantities by using the following identity: 
\begin{equation}
\int \bra{\psi} \E(E_m^\dagger P_\psi E_n)\ket{\psi} d\psi = \frac{D \chi_{mn}+\delta_{mn}}{D+1}.
\label{eq:offdiagintegral}
\end{equation}
This equation can be obtained by using the $\chi$--representation for the map $\E$, together with equation (\ref{eq:avgastraces}) and the trace preserving condition for $\E$:
\begin{equation}
 \sum_{m'n'} \chi_{m'n'} \trace{E_{m'}E_m^\dagger E_n E_{n'}^\dagger} = \trace{E_m^\dagger E_n} = D\delta_{m,n}
\end{equation}
To measure the complex off-diagonal coefficients $\chi_{mn}$ we can proceed as follows: We add an extra clean qubit and consider the map $\E_{mn}$ to be the one described by the circuit in the dashed box of Fig. 2.
\begin{figure}[h]
\begin{equation*}
\Qcircuit @C=.9em @R=.4em {
     \ket{0}_{\text{Ancilla}} & & \gate{H} &
         \ctrl{1} & \ctrlo{1} &
        \qw & \meter & \sigma_x
     \\
     \ket{\psi}_{\text{Main}} & & {/}\qw &   
         \gate{E_m^\dagger} & \gate{E_n^\dagger} & \gate{\mbox{\huge $\E$}} &
          \meter & ~P_\psi
\gategroup{1}{3}{2}{6}{.7em}{--}
}
\end{equation*}
\caption{Circuit for measuring $\Real{\chi_{mn}}$ for a given channel $\E$}
\label{circ:offdiag} 
\end{figure}
Thus, if $C$-$E_m$ denotes the controlled $E_m$ operation, we have $\E_{mn}(\sigma)=\E(C$-$E^\dagger_n \bar C$-$E^\dagger_m H \sigma H \bar C$-$E_m C$-$E_n)$ (where $\sigma$ is the joint state of the ancillary qubit and the original system).
After applying this operation we can use equation (\ref{eq:offdiagintegral}), and realize that the real part of $\chi_{mn}$ can be obtained by measuring the polarization of the ancillary qubit (i.e. the expectation value of $\sigma_x$ conditioned on the survival of the state $\ket{\psi}$, averaging over all states $\ket{\psi}$). Thus, this is due to the fact that 
\begin{equation}
\int \trace{\E_{mn}\left(\ket{0}\bra{0}\otimes P_\psi\right)\sigma_x\otimes P_\psi} d\psi=\frac{D\Real{\chi_{mn}}+\delta_{mn}}{D+1}, 
\end{equation}
If instead of measuring the expectation of $\sigma_x \otimes P_\psi$ we measure the expectation value of $\sigma_y \otimes P_\psi$, the average over $\ket{\psi}$ yields $\frac{D\Imag{\chi_{mn}}}{D+1}$.

Summarizing: So far we showed that all matrix elements $\chi_{mn}$ can be directly related to average survival probabilities. In fact, the diagonal coefficients $\chi_{mm}$ are related to average fidelities of the channel $\E_m$. The off diagonal coefficients are related to the average polarization of the ancillary qubit conditioned to the survival of the state after the evolution with the channel $\E_{mn}$. 

Experimentally measuring these averages over the Hilbert space seems completely unrealistic. However, the beautiful recent work on the theory of $2$--designs 
\cite{Dan05:MT, DCEL06:arxiv, Ambainis07, klappenecker-2005} provides the means for doing so. P. Delsarte \cite{Del77} showed how integrating polynomials on the sphere could be reduced to averaging the integrand on a finite set of points coined spherical designs. The same idea can be extended to integrals over the entire Hilbert space. Here, we only need a state $2$--design $K$, that satisfice
\begin{equation}
\int \bra{\psi}O_1 P_\psi O_2 \ket{\psi} d\psi = 
\frac{1}{\abs{X}}\sum_{\psi \in X} \bra{\psi}O_1 P_\psi 
O_2 \ket{\psi},
\label{eq:2design}
\end{equation}
for all operators $O_{1,2}$. Thus, averaging over the entire Hilbert space is equivalent to averaging over the finite set $X$. State $2$--design with a finite (but exponentially large) number of states exist. Although the computation of the exact average over the $2$--design is still exponentially hard, it is now possible to realize that an estimate for the average can be efficiently found. 
This average can be estimated by randomly sampling over initial states $\ket{\psi}$ chosen from the set $X$. 
This is the final piece of our method. 

Luckily, it is rather easy to produce a state $2$--design for $n$ qubits.
One possibility is finding $D+1$ mutually unbiased bases (MUB) that 
automatically form a state $2$--design \cite{klappenecker-2005}. 
Each base will be labeled with an index $J=0,\ldots,D$ and the states within each base will be labeled with the index $m=1,\ldots,D$. 
In order for the orthonormal bases to be unbiased, 
the $D(D+1)$ states of the MUBs must satisfy
$\abs{\braket{\psi^J_m}{\psi^{K}_{n}}}^2 = \frac{1}{D}$ for all $J\neq K$.
Since generalized Pauli operators may be partitioned into $D+1$ maximally sets of $D$ commuting operators so that each pair of sets only hold the identity $I$ as common element \cite{bandyopadhyay-new}, there are $D+1$ MUBs, each one diagonalizing each of these commuting subsets of Pauli operators \cite{lawrence-2002-65}. In this way, the set of states in the MUBs can be efficiently described and also can be efficiently generated with $O(n^2)$ one and two qubit gates \cite{Ben06:TF}. It is simple to adapt the procedure used to efficiently generate any state in any MUB to compute survival probabilities of such states and also to compute the transition rates from the $(J,m)$ to $(J,m')$ states. 

Other strategies may use $2$--designs other than MUBs. For example, Dankert et. al. \cite{Dan05:MT, DCEL06:arxiv} propose to use of approximate unitary $2$--designs (which are designs on the space of unitary operators) showing that they can be efficiently approximated. 
An approximate unitary $2$--design with accuracy $\epsilon + 1/D^2$ can be obtained by employing only $O(n \text{ log }\frac{1}{\epsilon})$ gates. 
Unitary $2$--designs acting on any fixed state induce state $2$--designs fitting into the previous scheme. 
Dually the action of the random unitaries may be interpreted as symmetrizing the channel $\E$ through twirling.
Following this line, we may also use weaker symmetrization protocols as in SCNQP \cite{JosephEmerson09282007} for estimating fidelities of modified channels (\ref{circ:diag},\ref{circ:offdiag}).

Let us summarize the complete method. 
The estimation of diagonal coefficients $\chi_{mm}$ of the quantum map $\E$ can be efficiently done by evaluating the fidelity of the channel $\E_m$  averaged over a random sample of the $2$--design formed by the states of $D+1$ MUBs. 
The real and imaginary parts of the off-diagonal elements $\chi_{mn}$ are obtained from the $x$ and $y$--polarizations of the ancillary qubit conditioned on the survival of the states randomly chosen from the same $2$--design when the evolution of the combined system+ancilla is described by the map $\E_{mn}$.
The strongest requirements for this method to be applicable are $2$--design state generation and the implementation of the single qubit controlled Pauli operators used in the construction of the $\E_{mn}$ channels.

The fact that the estimate of $\chi_{mn}$ can be obtained by sampling a small subset of states is crucial for the efficiency of the method. 
For definiteness, suppose we are interested in estimating  $\frac{D\Imag{\chi_{mn}}+1}{D+1}$ with precision $\epsilon$.
The results of a given experimental run may produce three outputs: $0$, corresponding to no survival on the main system, and $\pm 1$ corresponding to survival and different polarizations eigenstates on the auxiliary output.
Hence, independent measures have a variance smaller than $1$.
Averaging over $M \geq \epsilon^{-2} $ independent experiments, we  may obtain the desired precision. Similarly the estimation of fidelities $F(\E_m)$ may achieve a precision $\epsilon$ using only $M \geq \epsilon^{-2}/4$  independent experiments.
The important result is that the number of required experiments does not increase with $D$.

Up to this point we have only considered measuring survival probabilities.
However, after preparing a state $\ket{\psi^{J}_{k}}$, we can measure transition probabilities to other states $\ket{\psi^{J}_{k'}}$. We will show that in this way we can efficiently estimate all the diagonal coefficients $\chi_{mm}$.  
To do this, it is crucial to use the $2$--design formed by the MUBs associated to the same operators used in the $\chi$--matrix description of $\E$ (the Pauli operators are well suited for that purpose). Suppose that we perform $M$ experiments and we store the results as triplets $(J,k,k')$ corresponding to the labels of the initial state and the one detected. We will show that the fidelity $F(\E_m)$ can be estimated as the frequency of certain events (events of the m-type). Below, we will describe a simple procedure to decide if a given event is of the m-type or not (the procedure requires resources polynomial in $n$). 

Each base is determined by $n$ commuting Pauli operators $P^J_1 \ldots P^J_n$ which are the generators of the Abelian group of operators that are diagonal in that particular basis. We can label the base elements ($\ket{\psi^J_k}$) according to the eigenvalues of the $n$ generators. 
Thus, the label $k$ is an $n$ components binary vector: the $i$--th component $k_i$ determines the $\pm 1$ eigenvalue of $P^J_i$ as $P^J_i \ket{\psi^J_k}= (-1)^{k_i}\ket{\psi^J_k}$. The action of any Pauli $P$ on the state $\ket{\psi^J_k}$ transforms it into another state of the same base $J$. 
Those transition rules are fully determined by the binary vector $p$ that encodes the commutation pattern between $P$ and the generators of the $J$--th base (i.e. the vector $p$ is such that $(-1)^{p_i}P P^J_i = P^J_i P$ ).
Thus, $P\ket{\psi^J_k} \simeq \ket{\psi^J_{k+p}}$ (up to a phase), where $k$ and $p$ are added bitwise modulo $2$. 
Hence, the probability to detect the state $\ket{\psi^J_{k+p}}$ in the final measurement after evolving with any unitary operator $U$ is identical to the probability of detecting the state $\ket{\psi^J_k}$ after evolving with the operator obtained as the product $P\times U$. 
Applying this observation we conclude that in order to measure any diagonal coefficient $\chi_{mm}$ we could modify the strategy we described above by removing the final $E_m$ gate proceeding as follows: We compute $\chi_{mm}$ by averaging the transition probability to the state $\ket{\psi^J_{k+p_m}}$ whenever the state $\ket{\psi^J_k}$ is prepared ($p_m$ is the commutation vector of $E_m$ and the generators of the base $J$).

The procedure to estimate all diagonal $\chi_{mm}$ coefficients must be clear now. To estimate the associated fidelity $F(\E_m)$ from the experimental data obtained after $M$ experiments we must compute the frequency of the events $(J,k,k')$ that satisfy the condition $k+k'=p_m$ (where $p_m$ is the commutation vector of the operator $E_m$ and the generators of the base $J$). These are the m-type events. Performing this check requires knowing the vector $p_m$ (which requires $O(n^2)$ classical operations). 
The lassical complexity for this process is $O(n^2M)$. 
However, to the best of our knowledge, the construction of the $n$ generators for base $J$ requires $O(n^4)$ classical operations, which is the dominating overhead.
Errors corresponding to the estimations of the different $\chi_{mm}$ are correlated. However the variance of any such estimator behaves in the same way as the one corresponding to the method to evaluate a single $\chi_{mm}$ at a time. 

We can also use this method to devise a test to efficiently detecting the coefficients $\chi_{mm}$ with values that are above a certain threshold (which is $D$--independent, i.e. not exponentially small). The idea is that if the coefficient $\chi_{mm}$ is large then the condition $k+k'=p_m$ must be satisfied frequently in the experimental data. Given the triplets corresponding to two experiments performed on different bases, we may efficiently determine the unique operator $E_m$ satisfying the condition for both of them (using $O(n^3)$ operations). The number of such pairs of triplets is bounded by $M(M+1)/2$. 
Therefore we can find out the operators $E_m$ satisfying the condition $k+k'=p_m$  for at least two bases $J$ with an overhead that depends at most quadratically $M$. All the operators satisfying such criterion will be obtained with $O(M^2 n^3)$ classical operations. After sieving the operators in this way we can focus on those that passed the test to study them further. The conclusion is that in this way, it is possible to efficiently identify and estimate all large $\chi_{mm}$ coefficients. Again, we stress that this method enables us to estimate if the coefficients are larger than a fixed $D$-independent threshold such as $M^{-1/2}$. We may also simultaneously estimate groups of off diagonal $\chi_{mn}$ coefficients for which $E_m^\dagger E_n$ is the same. The required book-keeping is similar to the one used for diagonal coefficients but requires additional care regarding the numerical phases that arise.

\emph{Outlook.}-- We have shown how any particular $\chi_{mn}$ coefficient for a channel $\E$ may be estimated from the average survival probabilities of at most two modified channels requiring at most one extra clean qubit.
This approach, together with the use of randomized fidelity measurement schemes is the first to allow the efficient estimation of any desired $\chi_{mn}$ coefficient with a number of experiments that is dimension independent.
The method shows its strength when only partial tomography is desired, since the resources required are polynomial in the number of subsystems.
We further presented a possible extension by sampling over initial states belonging to the MUBs associated with the same operator base used in the channel representation. This allows us to profit from the information provided by complete base measurements. 
Finally, we prove the possibility of efficiently detecting and characterizing Pauli channels with \emph{sparse} $\chi$--matrices. JPP is a Fellow of CONICET. This work was supported with grants from Anpcyt and Conicet. 
\bibliography{esqpt}

\end{document}